\documentclass[twocolumn,showpacs,preprintnumbers,amsmath,amssymb]{revtex4}
\usepackage{graphicx}% Include figure files
\usepackage{dcolumn}% Align table columns on decimal point
\usepackage{bm}% bold math

\def\be{\begin{equation}}
\def\ee{\end{equation}}
\def\e#1{\label{#1}\end{equation}}
\def\bea{\begin{eqnarray}}
\def\eea{\end{eqnarray}}
\def\ea#1{\label{#1}\end{eqnarray}}

\def\bem#1{\begin{mathletters}\label{#1}}
\def\eml{\end{mathletters}}

\def\ket#1{{|#1\rangle}}
\def\bra#1{{\langle#1|}}
\def\braket#1#2{{\langle#1|#2\rangle}}
\def\4#1{{\boldsymbol{#1}}}
\def\8#1{{\widetilde{#1}}}

\begin{document}

\preprint{APS/123-QED}

\title{Preventing Multipartite Disentanglement by Local Modulations}

\author{G. Gordon}
\author{G. Kurizki}
\email{gershon.kurizki@weizmann.ac.il}
\affiliation{%
Department of Chemical Physics, Weizmann Institute of Science,
Rehovot 76100, Israel
}%

\date{\today}% It is always \today, today,
\begin{abstract}
An entangled multipartite system coupled to a zero-temperature
bath undergoes rapid disentanglement in many realistic scenarios,
due to local, symmetry-breaking, differences in the particle-bath
couplings. We show that locally controlled perturbations,
addressing each particle individually, can impose a symmetry, and
thus allow the existence of decoherence-free multipartite
entangled systems in zero-temperature environments.
\end{abstract}



\maketitle
Symmetry is a powerful means of protecting entangled
quantum states against decoherence \cite{aku05}, since it allows
the existence of a decoherence-free subspace or a decoherence-free
subsystem (DFS) \cite{zan97,vio00}. In multipartite systems, this
requires that all particles be perturbed by the {\em same}
environment. In keeping with this requirement, quantum
communication protocols based on entangled two-photon states have
been studied under {\em collective} depolarization conditions,
namely, {\em identical} random fluctuations of the polarization
for both photons \cite{ban04}. Entangled qubits that reside at the
same site or at equivalent sites of the system, e.g. atoms in
optical lattices, have likewise been assumed to undergo identical
decoherence \cite{aku05}.

However, locally-decohering entangled states of two or more
particles, such that each particle travels along a different
channel or is stored at a different site in the system, may break
this symmetry. A possible consequence of this symmetry breaking is
the abrupt ``death'' of the entanglement \cite{yu04}. Such
systems, composed of particles undergoing individual or ``local''
decoherence, do not possess a natural DFS and thus present more
challenging problems insofar as decoherence effects are concerned
\cite{moe02}. Specifically, can the multipartite symmetry, broken
by ``local'' decoherence, be dynamically restored to allow a DFS?
Alternatively, can we dynamically impose other symmetries that
protect {\em all particles}, not just part of them, from local
decoherence? Previous treatments \cite{vio00} have suggested to
dynamically restore a DFS by subjecting the particles to
sufficiently frequent, fast and strong pulses, assuming that they
stroboscopically decouple all particles from the bath,
irrespective of their coupling strengths.

Here we address these fundamental questions by developing a
generalized treatment of multipartite entangled states (MES)
decaying into zero-temperature baths and subject to {\em arbitrary}
external perturbations whose role is to provide {\em dynamical
protection} from decay and decoherence. Our treatment applies {\em
to any difference} between the couplings of individual particles to
the baths. It does not assume the perturbations to be stroboscopic,
i.e. strong or fast enough, but rather to act concurrently with the
particle-bath interactions. Our main results are to show that by
applying {\em local} (selective) perturbations to multilevel
particles, i.e. by {\em addressing each level and each particle
individually}, one can create a decoherence-free system of many
entangled qubits. Alternatively, one may reduce the problem of
locally decohering MES to that of a single decohering particle,
whose dynamical control has been thoroughly investigated
\cite{aga01,kof01,gor05}.

In our treatment we introduce the multipartite system and arrive at
a general dynamical solution for its decoherence. We then analyze
realizations of different symmetries of the decoherence matrix
comparing the resulting fidelities of the decohering state. Our
system comprises $N$ particles, labelled by $j=1,...,N$, each having
a ground state $\ket{g}_j$ and $M_j$ excited states, $\{\ket{n}_j\}$
with energies, $\hbar\omega_{j,n}$. In particular, $n_j$ may
enumerate the motional states of a trapped ion or atom,
\cite{moe02}. The coupling to the zero-temperature bath may differ
from one particle to another and from one excited level to another.
For their protection from decay and decoherence, we apply a
perturbation such that each level is modulated by {\em a different
AC Stark shift} $\hbar\delta_{j,n}(t)$ and/or by different
off-resonant perturbing field $\tilde\epsilon_{j,n}(t)$. The total
Hamiltonian is the sum of the multipartite system, bath and {\em
off-diagonal system-bath interaction} Hamiltonians \cite{scu97}:
\bea
\label{H-total} &&H(t)=\hbar\sum_{j=1}^N\sum_{n=1}^{M_j}
\left[\omega_{j,n}+\delta_{j,n}(t)\right] \ket{n}_j\,{}_j\bra{n}+
\hbar\sum_k\omega_k\ket{k}\bra{k}+ \nonumber\\
&&\hbar\sum_{j=1}^N\sum_k\sum_{n=1}^{M_j}[\tilde\epsilon_{j,n}(t)\mu_{k,j,n}
\left(\ket{n}\bra{g}\right)_j\ket{vac}\bra{k}\otimes I_{j'\neq
j}+H.c.]
\eea
Here $I$ is the identity operator, $\mu_{k,j,n}$ is the
off-diagonal coupling coefficient of the $n^{th}$ excited level of
particle $j$ to an excitation of the bath mode $\ket{k}$ and
$\ket{vac}$ is the vacuum state of the bath. $H.c.$ are Hermitian
conjugates.

Although our strategy applies in general to any number of
excitations, simple closed-form solutions are obtainable for a
{\em single} initial excitation of the system. Accordingly, the
complete wave function is:
\be
\label{zero-gen-full}
\ket{\Psi(t)}=\sum_k\alpha_0^k(t)\ket{k}\bigotimes_{j=1}^N\ket{g}_j
+\sum_{j=1}^N\sum_{n=1}^{M_j}\alpha_{j,n}(t)\ket{n}_j\ket{vac}
\bigotimes_{j'\neq j}\ket{g}_{j'}
\ee
We will denote the first and second RHS terms as the bath and
system wave function, $\ket{\Psi^B(t)}$ and $\ket{\Psi^S(t)}$,
respectively. In order to solve the Schr\"{o}dinger equation, one
may eliminate the $\{\alpha^k_0(t)\}$ amplitudes and transform to
the interaction picture, ending up with an {\em exact}
integro-differential equation. Assuming that these amplitudes are
slowly varying on the time-scale of the bath response \cite{kof01}
and using the matrix representation, this equation has the general
solution:
\be
\label{alpha-def}
\tilde{\4\alpha}(t)=\mathrm{T_+}e^{-\4J(t')}\tilde{\4\alpha}(0),
\quad \tilde{\alpha}_{j,n}=
e^{i\omega_{j,n}t+i\int_0^td\tau\delta_{j,n}(\tau)}\alpha_{j,n}.\\
\ee
where $\tilde{\4\alpha}=\{\tilde\alpha_{j,n}\}$, $\mathrm{T_+}$ is
the time-ordering operator and $\4J(t)=\{J_{jj',nn'}(t)\}$ is the
dynamically-modified decoherence matrix, determined by the
following convolution:
\bea
\label{zero-gen-J-def} &&J_{jj',nn'}(t) = 2\pi
\int_{-\infty}^\infty d\omega
G_{jj',nn'}(\omega)K_{t,jj',nn'}(\omega)\\
&&G_{jj',nn'}(\omega)=\hbar^{-2}\sum_k\mu_{k,j,n}\mu^*_{k,j',n'}
\delta(\omega-\omega_k)\\
\label{K-def} &&K_{t,jj',n''}(\omega)=
\epsilon^*_{t,j,n}(\omega-\omega_{j,n})
\epsilon_{t,j',n'}(\omega-\omega_{j'n'})\\
&&\epsilon_{t,j,n}(\omega)=\int_0^td\tau\epsilon_{j,n}(\tau)e^{i\omega\tau}\\
\label{epsilon-tjn}
&&\epsilon_{j,n}(t)=\tilde\epsilon_{j,n}(t)e^{-i\int_0^td\tau\delta_{j,n}(\tau)}
\eea
Here $G_{jj',nn'}(\omega)$ is the coupling spectrum matrix given by
nature and $K_{t,jj',nn'}(\omega)$ is the dynamical modulation
matrix, which we design at will to suppress the decoherence (cf.
\eqref{H-total}). This general solution holds {\em for
dynamically-modified relaxation of a singly-excited MES into a
zero-temperature bath.}

For example, we may control the decoherence by impulsive phase
modulation, i.e. a sequence of pulsed Stark shifts caused by
fields whose amplitudes satisfy \cite{kof01}
\bea
\label{epsilon-t}
&&\epsilon_{j,n}(t)=e^{i[t/\tau_{j,n}]\theta_{j,n}}\\
\label{epsilon-omega}
&&\epsilon_{t,j,n}(\omega)=\frac{\left(e^{i\omega\tau_{j,n}}-1\right)
\left(e^{i(\theta_{j,n}+\omega\tau_{j,n})[t/\tau_{j,n}]}-1\right)}
{i\omega\left(e^{i(\theta_{j,n}+\omega\tau_{j,n})}-1\right)}.
\eea
Here $[...]$ denote the integer part, $\tau_{j,n}$ and
$\theta_{j,n}$ are the pulse duration and the phase change for
level $n$ of particle $j$, respectively. In the limit of weak
pulses, of area $|\theta_{j,n}|\ll\pi$, Eq.~\eqref{epsilon-omega}
yields
$\epsilon_{t,j,n}(\omega)\cong\epsilon_{t,j,n}\delta(\omega-\Delta_{j,n})$,
where $\Delta_{j,n}=\theta_{j,n}/\tau_{j,n}$ is the effective
spectral shift caused by the pulses.

It is expedient to rewrite the fidelity of the evolving system,
$F(t) = |\braket{\Psi^S(0)}{\Psi^S(t)}|^2$, as a product of two
factors:
\bea
\label{fidelity}
&&F(t)=F_p(t)F_c(t)\\
&&F_p(t)=|A(t)|^2=\sum_{j=1}^N\sum_{n=1}^{M_j}|\alpha_{j,n}(t)|^2\\
&&F_c(t)=\frac{|\sum_{j=1}^N\sum_{n=1}^{M_j}\alpha_{j,n}^*(0)\alpha_{j,n}(t)|^2}
{|A(t)|^2}
\eea
where $F_c(t)$ is the autocorrelation function,
$|\braket{\Psi^S(0)}{\Psi^S(t)}|^2$, normalized by the total
excitation probability $|A(t)|^2$. Thus $1-F_p(t)$ measures
population loss from any $\ket{n}_j$, whereas $1-F_c(t)$ is a
measure of correlation preservation: $F_c=1$ when the initial
multipartite correlations are completely preserved. As shown below
population and correlation preservation can be {\em independently
controlled} in the model of
Eqs.~\eqref{H-total},\eqref{zero-gen-full}.

In the absence of dynamical control, $F_c(t)$ decays much faster
than $F_p(t)$ and is much more sensitive to the asymmetry between
local particle-bath couplings. Thus, for initial Bell singlet and
triplet states,
$\ket{\Psi(0)}=1/\sqrt{2}(\ket{g}_A\ket{n}_B\pm\ket{n}_a\ket{g}_B)$,
which do not experience cross-decoherence but only different local
decoherence rates, $\alpha_{A(B)}(t)=1/\sqrt{2}e^{-J_{A(B)}}(t)$.
We find $F_p(t)=(e^{-2J_A(t)}+e^{-2J_B(t)})/2$;
$F_c(t)=(1+C(t))/2=1/2+e^{-\Delta J(t)}/(1+e^{-2\Delta J(t)})$,
where $\Delta J(t) = J_A(t)-J_B(t)$ and $C(t)$ is the concurrence
\cite{woo98}.

We shall first deal with $N$ identical qubits, and thus ignore the
$n$ subscript, i.e. set $\omega_j\equiv\omega_0$. We also require
that at any chosen time $t=T$, the AC Stark shifts in
Eq.~\eqref{alpha-def} satisfy, $\int_0^Td\tau\delta_j(\tau)=2\pi
m$, where $m=0,\pm1,...$. This requirement ensures that
modulations only affect the decoherence matrix
\eqref{zero-gen-J-def}, but do not change the relative phases of
the entangled qubits when their MES is probed or manipulated by
logic operations at $t=T$.

\begin{figure}[ht]
\centering\includegraphics[width=8.5cm]{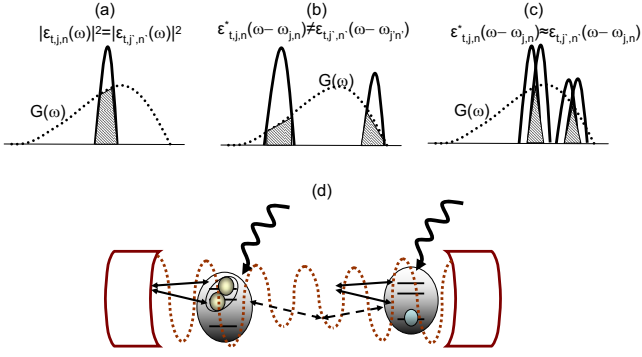}
 \caption{Two 3 - level particles in a cavity, coupled to the cavity
 modes (thin lines) and subject to local control fields (thick lines). (a-c) Frequency domain
 overlap of coupling spectrum (dotted) and modulation matrix elements(solid),
 resulting in modified decoherence matrix elements (shaded), for:
 (a) global modulation, (b) cross-decoherence elimination (IIP symmetry) and (c) IIT symmetry.
 (d) IIT symmetry scheme.}
 \protect\label{Fig-1}
\end{figure}

Without any modulations, decoherence in this scenario has no
inherent symmetry. Our point is that one can symmetrize the
decoherence by appropriate modulations. The key is that different,
``local'', phase-locked modulations applied to the individual
particles, according to Eq.~\eqref{K-def}, can be chosen to cause
{\em controlled interference} and/or spectral shifts between the
particles' couplings to the bath. The $K_{t,jj'}(\omega)$ matrices
(cf.\eqref{K-def}) can then satisfy $2N$ requirements at all times
and be tailored to impose the advantageous symmetries described
below. By contrast, a ``global'' (identical) modulation,
characterized by $K_{t,jj'}(\omega)=|\epsilon_t(\omega)|^2$, is
not guaranteed to satisfy $N\gg1$ symmetrizing requirements at all
times (Fig.~\ref{Fig-1}a).
The most desirable symmetry is that of {\em identical coupled
particles} (ICP), which would emerge if all the modulated
particles could acquire the {\em same} dynamically modified
decoherence and cross-decoherence yielding the following $N\times
N$ fully symmetrized decoherence matrix
\be
\label{J-ICP}
J_{jj'}^{\rm ICP}(t)=r(t)\quad\forall j,j'.
\ee
ICP would then give rise to a $(N-1)$-dimensional decoherence-free
subspace: the entire single-excitation sector less the totally
symmetric entangled state. An initial state in this DFS
\cite{zan97} would have $F(t)=1$ for all times, meaning that it
would neither lose its population nor its initial correlations (or
entanglement).

However, it is generally impossible to ensure this symmetry, since
it amounts to satisfying $N(N-1)/2$ conditions using $N$
modulating fields. Even if we accidently succeed with $N$
particles, the success is not scalable to $N+1$ or more particles.
Moreover, the ability to impose the ICP symmetry by local
modulation fails completely if not all particles are coupled to
all other particles through the bath, i.e. if some
$G_{jj'}(\omega)$ elements vanish.

\begin{figure}[ht]
\centering\includegraphics[width=8.5cm]{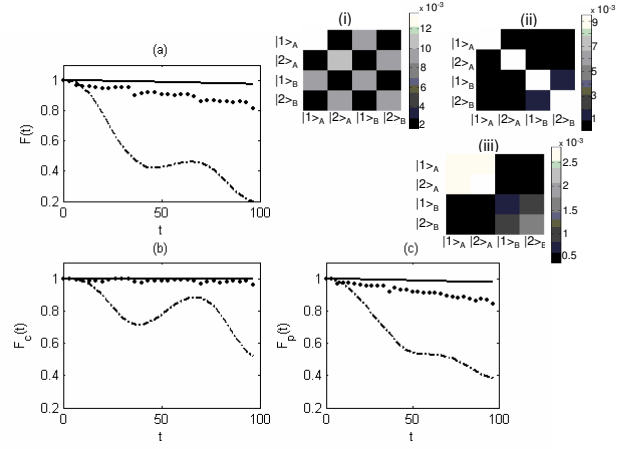}
 \caption{Fidelity as a function of time (in units of $(10\gamma)^{-1}$): (a) overall fidelity,
 (b) correlation preservation, and (c) population preservation.
 (i) Global $\pi$-phase flips impose no symmetry (dashed)
 ($\tau_{j,n}=1.1, \theta_{j,n}/\pi=1.0$). (ii) independent identical particles (IIP) symmetry (dotted)
 ($\tau_{j,n}=(0.75,0.85,0.95,1.05), \theta_{j,n}/\pi=(0.834,0.806,0.836,0.82)$);
 (iii) independent identical trapping (IIT) symmetry (solid)
 ($\tau_{j,n}=(0.85,0.85,1.05,1.05), \theta_{j,n}/\pi=(0.924,0.9,0.945,0.91)$).
 Top-right: Decoherence matrix elements at time $t=100$.
 The parameters are: $\omega_1=0.5, \omega_2=0.6$, $\gamma=0.1$, $k_0r_{min}=1.0$,
 $t_{j,n}=(0.7, 1.0, 1.06, 1.1)$,
 $r_j = (0.0, 0.1)$, $\eta_n/\pi=(0.0, 0.1)$. }
 \protect\label{Fig-2}
\end{figure}

A more limited symmetry that we may {\em ensure} for $N$ qubits is
that of {\em independent identical particles} (IIP). This symmetry
is formed when spectral shifts and/or interferences imposed by $N$
modulations cause the $N$ different particles to acquire the {\em
same} single-particle decoherence $r(t)$ and experience no
cross-decoherence. To this end, we may choose in
Eq.~\eqref{epsilon-t}
$\epsilon_{t,j}(\omega)\simeq\epsilon_{t,j}\delta(\omega-\Delta_j)$.
The spectral shifts $\Delta_j$ can be different enough to couple
each particle to a different spectral range of bath modes so that
their cross-coupling vanishes:
\be
\label{no-cross}
J_{jj'}(t)=\epsilon^*_{t,j}\epsilon_{t,j'}\int
d\omega
G(\omega)\delta(\omega-\Delta_j)\delta(\omega-\Delta_{j'})\rightarrow0.
\ee
Here, the vanishing of $G_{jj'}(\omega)$ for some $j,j'$ is not a
limitation. The $N$ single-particle decoherence rates can be
equated by an appropriate choice of $N$ parameters $\{\Delta_j\}$:
\be
J_{jj'}^{\rm
IIP}(t)=|\epsilon_{t,j}|^2G_{jj}(\Delta_j)=\delta_{jj'}r(t),
\ee
where $\delta_{jj'}$ is Kronecker's delta (Fig.~\ref{Fig-1}b). The
IIP symmetry results in complete correlation preservation, i.e.
$F_c(t)=1$, but still permits population loss,
$F(t)=F_p(t)=e^{-2{\rm Re} r(t)}$. If the single-particle $r(t)$
may be dynamically suppressed, i.e. if the spectrally shifted bath
response $G_{jj}(\omega_j+\Delta_j)$ is small enough, this $F(t)$
will be kept close to $1$ (Fig.~\ref{Fig-2}).
If IIP symmetry is imposed, the particles become effectively
independent in terms of their coupling to the bath. Hence,
collective coupling to the bath, which is a prerequisite for a DFS,
{\em is not formed} in this case. In order to impose
decoherence-free conditions, we extend the treatment to {\em
multilevel} particles \cite{gor05}. We will show that the use of
auxiliary levels and local modulations can {\em result in a
decoherence-free singly-excited $N$ qubit} system, in a general and
realistic coupling scenario.

To this end we invoke three-level particles with excited states
$\ket{1}$ and $\ket{2}$. In a single three-level particle, an
external field can impose an intraparticle DFS if $\ket{1}$ and
$\ket{2}$ decay at the same rate to the ground state $\ket{0}$;
this DFS consists of $\ket{0}$ and the ``dark'' or trapping state
\cite{scu97}, the anti-symmetric superposition of the two excited
states $1/\sqrt{2}(\ket{1}-\ket{2})$. However, this intraparticle
DFS would {\em be destroyed} in a system of $N$ three-level
particles that are coupled to the bath, and/or experience
cross-decoherence. In order to remedy this problem, let us
consider local modulations acting differently on the two excited
levels within each particle. Such modulations can be tailored to
impose what may be dubbed {\em ``independent identical trapping''}
(IIT) symmetry. This means that all particles acquire identical
trapping states and become (effectively) independent, {\em without
cross-decoherence}. As discussed above, $N$ spectral shifts
$\Delta_j$ determined by $|\epsilon_{t,j}(\omega)|^2$ suffice to
eliminate cross-decoherence, Eq.~\eqref{no-cross}. Under these
conditions, $N$ destructively interfering pairs of local fields
will cause trapping in each particle (Fig.~\ref{Fig-1}c,
\ref{Fig-1}d). This would result in an $N$-dimensional DFS,
composed of $N$ particles sharing single-excitation with
anti-symmetrically superposed excited states. Under the IIT
symmetry the decoherence matrix is block diagonal, each block
corresponding to particle $j$ with levels $n,n'$:
\be
\label{J-IIT} J_{jj',nn'}^{\rm IIT}(t)=\delta_{jj'}r_j(t)
\ee
Both disentanglement and population loss are nearly completely
eliminated within the multipartite subspace, resulting in $F(t)$
very close to $1$ (Fig.~\ref{Fig-2}).

As an example, the IIT recipe will be analyzed for two three-level
particles, where each level of each particle experiences different
coupling to the bath, and cross-decoherence exists. Let us take
the initial state of particles $A,B$ to be
$\ket{\Psi_S(0)}=\left(\ket{-}_A\ket{0}_B\pm\ket{0}_A\ket{-}_B\right)/\sqrt{2}$,
where $(j=A,B)$
$\ket{-}_j=\left(\ket{1}_j-\ket{2}_j\right)/\sqrt{2}$. The goal
will be to keep this state intact,
$\ket{\Psi_S(t)}\simeq\ket{\Psi_S(0)}$, by preventing
cross-decoherence and imposing intraparticle destructive
interference. The bath response matrix will be taken to be
$\Phi_{jj',nn'}(t)=\gamma d_nd_{n'}
\frac{e^{-t^2/4t_{j,n}^2}e^{-t^2/4t_{j',n'}^2}}
{k_0(r_{min}+r_{jj'})}$ where $\gamma$ is the single-particle
unperturbed decay rate (satisfying Fermi's Golden Rule
\cite{scu97}), $d_n=\cos\eta_n$, with $\eta_n$ being the angle of
transition dipole, $t_{j,n}$ is the correlation time of level $n$
of particle $j$, $k_0$ is the wavevector of the resonant
transition $\ket{0}\leftrightarrow\ket{1(2)}$, $r_{min}$ is the
distance below which the particles are identically coupled to the
bath and $r_{jj'}=|\4r_j-\4r_{j'}|$, where $\4r_j$ is the position
of particle $j$. This model may describe the {\em
distance-dependent cross-decoherence} of either radiatively or
vibrationally relaxing atoms or ions at different sites in a trap,
lattice or cavity \cite{fol05,kre04}. The parameters of the pulse
sequence in Eq.~\eqref{epsilon-t} for creating the IIP or IIT
symmetries were chosen such that the faster decaying particle
experienced the stronger and/or faster pulses, i.e.
$\theta_{j,n}/\tau_{j,n}$ were adapted to the bath response
parameters $\Phi_{jj,nn}$ of each particle. The pulse sequences
were chosen to be different enough for each particle in order to
eliminate the desired cross-decoherence terms, as per
Eq.~\eqref{no-cross}. Individual pulses were selected to obey the
other IIP or IIT requirements described above. The very high
fidelity achievable by IIT is seen in Fig.~\ref{Fig-2}.

We may apply these considerations, for example, to cold ions or
atoms in a cavity \cite{kre04}, whose radiative decoherence rates
depend on their position within the cavity and may vary by as much
as 15\%. This implies that placing several ions in the same cavity
breaks their multipartite symmetry. AC Stark shifts are then an
effective tool to restore the symmetry. To impose the IIP symmetry,
one requires an impulsive phase modulation at a rate exceeding the
cavity-mode linewidth, $1/\tau\geq\Gamma$ (typically $\sim12 GHz$).
In order to impose the IIT symmetry, one can either use two excited
states of the ion, or two ions for each qubit, such that their
singlet Bell state and the ground state form the DFS \cite{pet02}.
The modulations should be local, addressing each level and each ion
separately, and adjusting the rate and phase of the impulsive phase
modulation so as to impose the IIT multipartite symmetry, thus
creating an $N$-qubit DFS. Recent experiments \cite{kre04} indicate
that such addressability is feasible either by spatial or spectral
resolution.

To conclude, we have shown that local modulations are generally far
more apt than {\em identical (global) modulation to impose
multipartite symmetry on an otherwise completely asymmetrically
relaxing system}. In particular, local modulations can impose the
IIT symmetry, in a system of $N$ three-level particles, thus
creating a {\em decoherence-free singly-excited $N$-qubit system}.
The general formalism presented here spans the entire range of
possible coupling scenarios and modulation schemes. These results
imply that the challenge of multipartite decoherence, plaguing
quantum information transmission and storage, may be successfully
met in a variety of experimental situations.

We acknowledge the support of ISF and EC (QUACS and SCALA
Networks).

\end{document}